\documentclass[a4paper,11pt]{article}
\pdfoutput=1 

\usepackage{jinstpub} 

\usepackage{comment}
\usepackage{xcolor}

\newcommand{\nVHz}{${\textrm nV}/\sqrt{\textrm Hz}$}
\newcommand{\pAHz}{${\textrm pA}/\sqrt{\textrm Hz}$}

\title{\boldmath A low noise and low power cryogenic amplifier for single photoelectron sensitivity with large arrays of SiPMs}

\author[a]{P. Carniti,}
\author[a]{A. Falcone,}
\author[a]{C. Gotti\note{Corresponding author.},}
\author[b]{A. Lucchini,}
\author[a]{G. Pessina,}
\author[b]{S. Riboldi}
\author[a]{and F. Terranova}
\affiliation[a]{INFN Milano-Bicocca and University of Milano-Bicocca, Department of Physics\\Piazza della Scienza 3, Milano, 20126, Italy}
\affiliation[b]{INFN Milano and University of Milano, Department of Physics\\Via Celoria 16, Milano, 20133, Italy}

\emailAdd{claudio.gotti@mib.infn.it}

\abstract{This paper presents a low noise amplifier for large arrays of silicon photomultipliers (SiPMs) operated in cryogenic environments, especially liquid argon (87~K) and liquid nitrogen (77~K).
The goal is for one amplifier to read out a total photosensitive surface of tens of cm$^2$ while retaining the capability to resolve single photoelectron signals.
Due to the large capacitance of SiPMs, typically a few nF per cm$^2$, the main contributor to noise is the series (voltage) component.
A silicon-germanium heterojunction bipolar transistor (HBT) was selected as the input device of the cryogenic amplifier, followed by a fully differential operational amplifier, operated in an unconventional feedback configuration.
The input referred voltage noise of the circuit at 77~K is just below 0.4~\nVHz\ white (above 100~kHz) and 1~\nVHz\ at 10~kHz.
The value of the base spreading resistance of the HBT at 77~K was determined from noise measurements at different bias currents.
Power consumption of the full circuit is about 2.5~mW.
The design gives the flexibility to optimally compensate the feedback loop for different values of the input capacitance, and obtain a gain-bandwidth product in the GHz range.
The signal-to-noise ratio obtained in reading out SiPMs is discussed for the case of a 300~kHz low pass filter and compared with the upper limit that would derive from applying optimum filtering algorithms.
}

\keywords{Front-end electronics for detector readout, analogue electronic circuits, Photon detectors for UV, visible and IR photons (solid-state), Noble liquid detectors (scintillation, ionization, double-phase).}

\begin{document}
\maketitle
\flushbottom

\section{Readout of large arrays of SiPMs}
\label{sec:intro}

In recent years, silicon photomultipliers (SiPMs) emerged as a viable alternative to vacuum-based photomultipliers (PMTs) to sense scintillation and Cherenkov light signals in many kinds of particle detectors, since they offer similar or higher efficiency in a smaller and more robust package, and are not affected by magnetic fields.
Their main drawback is the significantly higher dark count rate (DCR), which, however, can be effectively mitigated by lowering the operating temperature.
For detectors that use liquid argon as a scintillator (such as the photon detection system of DUNE \cite{DUNE2016} and the DarkSide experiment \cite{DarkSide2018}), it is natural to take advantage of the cryogenic environment and operate the SiPMs inside the liquid (87~K).
At such temperature, dark counts of thermal origin become negligible, and those originated by tunneling dominate the DCR.
Proper shaping of the electric field and high silicon purity allow to reach DCR at the level of 0.01~Hz/mm$^2$ at cryogenic temperatures \cite{Collazuol2011, Acerbi2017}.
By connecting several SiPMs side by side, photosensitive surfaces of tens of cm$^2$ can then be realised, capable of resolving faint light signals down to single photons \cite{Dincecco2018}.

From the electrical point of view, with an approximation that will suffice for the rest of this paper, the source impedance of PMTs and SiPMs is capacitive.
But while for large area PMTs it is just the parasitic capacitance of the readout electrode and its connections, typically of the order of 10~pF or less, the source capacitance of SiPMs is given by the total capacitance of the cells it is composed of, and is then directly proportional to the photosensitive surface, with typical values of about 50~pF/mm$^2$.
When several SiPMs are connected in parallel (ganged) to instrument an area of tens of cm$^2$, a source capacitance of tens or hundreds of nF is to be expected.
Despite having similar gain and signal characteristics, large arrays of SiPMs have then a significantly lower source impedance than large area PMTs.
The importance of the parallel (current) noise of the amplifier is reduced, while its series (voltage) noise becomes the leading contributor.
Hence the need for a front-end amplifier designed for the lowest possible series noise.

The quest for low series noise is balanced by the need for low power consumption.
If the SiPMs and the amplifier are submerged in a cryogenic liquid, boiling should be avoided, since it would interfere with the propagation of photons and their detection by the SiPMs.
Another useful feature is the ability to drive the (differential) transmission lines that connect the output of the circuit to the outside world, i.e. the data acquisition systems located at room temperature, up to several meters away.

\section{Description of the circuit}

\begin{figure}[t]
\centering
\includegraphics[width=\textwidth]{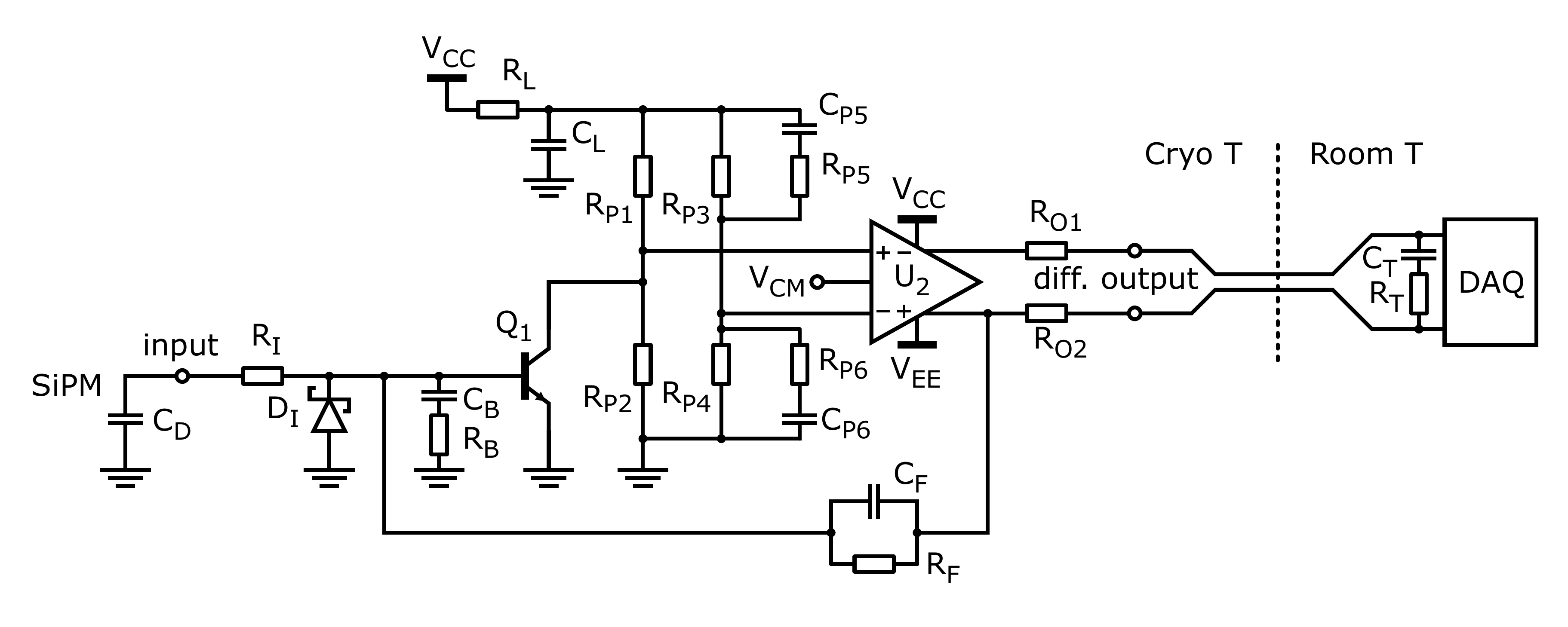}
\caption{\label{fig:schematic} Schematic of the amplifier. $Q_1$ is a silicon-germanium heterojunction bipolar transistor (Infineon BFP640). $U_2$ is a fully differential operational amplifier (Texas Instruments THS4531). A typical choice of component values is summarized in table \ref{tab:components}.}
\end{figure}

Figure \ref{fig:schematic} shows the schematic of the amplifier.
The SiPM is a 2-terminal device: the choice of anode or cathode readout affects the polarity of the bias voltage and of the current signals.
The input of the circuit is connected to the anode (cathode) of the SiPM, modeled by the source capacitance $C_D$.
The cathode (anode) of the SiPM, not shown, is connected to a bias voltage generator, bypassed to ground with a large value capacitor close to the SiPM.
The SiPM connected at the input is modeled by its capacitance $C_D$, expected to range up to $\sim$100~nF for a $\sim$20~cm$^2$ photosensitive area.

The circuit is based on a general and well known topology: a discrete transistor $Q_1$ followed by the operational amplifier $U_2$, although the choice of a fully differential opamp results in an unconventional feedback configuration, to be discussed in the following.
$Q_1$ is a silicon-germanium heterojunction bipolar transistor (HBT), designed for radiofrequency applications.
The presence of germanium atoms in the base makes its band-gap smaller than that of the emitter, allows higher doping levels, and results in a very small base spreading resistance, hence low series noise, and wide bandwidth even at low bias currents (below 1 mA).
These characteristics also make most HBTs able to work effectively at cryogenic temperatures, although in some cases with higher low frequency noise \cite{Joseph1995, Arnaboldi2003, Cressler2005, Bardin2008}.
Several cryogenic amplifiers are reported in literature that take full advantage of SiGe HBTs \cite{Kiviranta2006, Weinreb2009, Ivanov2011, Beev2012, Beev2013, Montazeri2016, Ying2017, Dark2019}.
The HBT we used in this work is the BFP640 from Infineon.

Due to the large value of $C_D$, abrupt changes in the bias voltage of the SiPMs could easily propagate to the input of the amplifier.
The Schottky diode $D_I$ protects $Q_1$ by guarding against reverse bias of its base-emitter junction.

The second stage $U_2$ is a fully differential operational amplifier.
The device we used as $U_2$ is the THS4531 from Texas Instruments, with 27~MHz differential gain-bandwidth product, 0.25~mA supply current and rail to rail outputs at room temperature.
Being designed in a BiCMOS technology, its ability to operate at 77~K is not to be taken for granted \cite{Lengeler1974}.
Band-gap narrowing, which can take place in the highly doped emitter of standard (homojunction) bipolar transistors, could degrade the current gain at low temperature and impair the performance.
MOS transistors, on the contrary, are generally able to work in cryogenic environments.
But even for a fully CMOS opamp, the ability to work at 77~K might also depend on other parameters related to circuit design (stability of voltage or current references, etc.), and needs in any case to be tested.
Several samples of the THS4531 were tested in liquid nitrogen, and were all observed to work, with supply current increased by about 50\%, and bandwidth almost doubled, reaching a gain-bandwidth product of 50~MHz.

The outputs of the circuit are connected to a data acquisition system (DAQ) through a differential transmission line of characteristic impedance $R_T$ or, equivalently, two lines of characteristic impedance $R_T/2$.
The termination is AC-coupled through $C_T$ to avoid DC current flowing in the output lines.
When the outputs of the THS4531 are terminated with high impedance (1~M$\Omega$), the output dynamic range at 77~K is still almost rail to rail, although a small oscillation with a frequency of about 20~MHz was observed.
The frequency of the oscillation does not appear to depend on the loop gain of the amplifier. It could be related with the compensation of the output stages of the opamp, which operate at unity gain. A complete study of this behaviour cannot be performed without a detailed description of the THS4531. 
The oscillation disappears when the outputs are AC-terminated with 50~$\Omega$, since in this case the loop gain of the rail-to-rail output stage is reduced; but in this case we observed the output dynamic range to be limited to about $\pm 1 V$.
The best compromise was then to couple the 50~$\Omega$ termination above a few MHz, by choosing $R_T=$100~$\Omega$, $C_T=$330~pF.
This proved to be effective in suppressing the aforementioned oscillation and preventing reflections in the output lines even when long cables (up to 12~m) were used, while at the same time limiting the load at the output of the amplifier, allowing it to work with the wide dynamic range it shows on a high impedance load.
This remains true as long as the timescale of the signals is larger than their propagation time on the output lines, of the order of tens of ns.

\begin{table}[b]

\centering
\caption{\label{tab:components} Choice of component values for the schematic of figure \ref{fig:schematic}.
With $V_{CC} = 3$~V, $V_{EE} = -1$~V, the input transistor is biased with $I_{C} = 370$~$\mu$A.
The ``-'' stands for ``not present''.
Unless otherwise noted, these are the values used in the measurements presented in this paper.}
\smallskip
\vspace{10pt}

\begin{minipage}{.3\linewidth}

\centering
\begin{tabular}{|r|l|}
\hline
$Q_1$ & BFP640 \\
$U_2$ & THS4531 \\
$D_I$ & SB01-15C \\
$R_I$ & 1 $\Omega$ \\
$R_F$ & 1.2 k$\Omega$ \\
$C_F$ & - \\
$R_B$ & - \\
$C_B$ & - \\
$R_{O1}$ & 50 $\Omega$ \\
$R_{O2}$ & 50 $\Omega$ \\
\hline
\end{tabular}

\end{minipage}%
\begin{minipage}{.3\linewidth}

\centering
\begin{tabular}{|r|l|}
\hline
$R_{L}$ & 510 $\Omega$ \\
$C_{L}$ & 100 nF\\
$R_{P1}$ & 3 k$\Omega$ \\
$R_{P2}$ & 16 k$\Omega$\\
$R_{P3}$ & 16 k$\Omega$\\
$R_{P4}$ & 16 k$\Omega$\\
$R_{P5}$ & 3.3 k$\Omega$ $+$ 390 $\Omega$\\
$C_{P5}$ & 100 nF\\
$R_{P6}$ & 100 $\Omega$\\
$C_{P6}$ & 2.2 nF\\
\hline
\end{tabular}

\end{minipage}%
\end{table}

The signal from the SiPM can be modeled as a current pulse $I_D (t)$ with a fast rise ($\leq$10~ns) and a slower recovery ($\tau_D \sim$~100~ns or larger).
Let us now assume $U_2$ to be ideal, with infinite open-loop gain and bandwidth, and let us neglect $R_I$.
The gain block made of $Q_1$ and $U_2$ can be seen as a single high-gain opamp with (negative) feedback provided by $R_F$ and $C_F$.
If $C_F$ is small enough ($C_F R_F < \tau_D$), there is no significant integration of the SiPM signals, and the transimpedance closed loop gain is equal to $R_F$.
Similarly, $R_I$ can be considered ``small enough'' if there is no integration at the input node, that is for $C_D R_I < \tau_D$.
Under these assumptions, the differential signal across the outputs is simply given by $V_O (t)= - 2 R_F I_D (t)$.
In most of the measurements presented in this paper, we worked without $C_F$, and with $R_I = 1$~$\Omega$, so the assumption to neglect them here is indeed justified.

The DC voltage at the non-inverting output of $U_2$ is the $V_{BE}$ of $Q_1$, around 0.6~V at room temperature, 1~V at 77~K.
The other output of $U_2$ is at $V_{CM} - V_{BE}$, where $V_{CM}$ is the voltage applied at the common-mode input of $U_2$.
$V_{CM}$ can be chosen to maximize the output dynamic range, the natural choice being halfway between the power supplies of $U_2$.
The purpose of $R_L$, $C_L$ is to filter high frequency noise and disturbances from $V_{CC}$. Let us neglect the voltage drop across $R_L$.
At DC, the inverting input of $U_2$ is held at a fixed voltage by the divider made by $R_{P3}$ and  $R_{P4}$. When the feedback loop is closed, this sets the voltage at the collector of $Q_1$.
If $R_{P3}=R_{P4}$, and neglecting $R_L$, the collector of $Q_1$ is at $V_{CC}/2$. The current through $R_{P1}$ is then ${V_{CC}}/{2 R_{P1}}$, while the bias current of $Q_1$ is
\begin{equation}
I_{C} = \frac{V_{CC}}{2}\left(\frac{1}{R_{P1}}-\frac{1}{R_{P2}}\right)
\end{equation}
where clearly $R_{P2}$ needs to be larger than $R_{P1}$.
Table \ref{tab:components} lists the typical values of the components used in the measurements throughout the paper, unless specified otherwise.
With $V_{CC} = 3$~V, $R_{P1} = $~3~k$\Omega$, $R_{P2} = R_{P3} = R_{P4} = $~16~k$\Omega$, the $V_{CE}$ of $Q_1$ is 1.5~V and its bias current is approximately 400~$\mu$A.
Since $R_L=510$~$\Omega$ gives a voltage drop of about 200~mV, $I_{C}$ is actually 370~$\mu$A.

Let us neglect $R_{P6}$ and $C_{P6}$ for now.
The role of $R_{P2}$ and $R_{P5}$, which is AC coupled through $C_{P5}$, is to make the inputs of $U_2$ as symmetrical as possible.
In particular, $R_{P2} = R_{P4}$, while $R_{P5}$ is chosen so that $R_{P3} \parallel R_{P5} = R_{P1}$.
$C_{P5}$ is as large as possible, compatibly with the requirement to work at cryogenic temperature, which led us to choose C0G (NP0) ceramic capacitors.
In case larger values are needed, solid tantalum capacitors could also be used \cite{Hammoud1998}.
This arrangement improves substantially the capability of the circuit to reject disturbances from the power supply $V_{CC}$, which is then limited only by the resistor precision (typically to about 40~dB for 1\% resistors).
With $C_{P5}=$~100~nF, $R_{P5} = 3.69$~k$\Omega$ (obtained as $3.3$~k$\Omega$ in series with $390$~$\Omega$), the rejection is effective starting from a few hundred Hz.
At high frequency, above about 1 MHz, the inverting input of $U_2$ goes to ground through $R_{P6}=100$~$\Omega$ and $C_{P6}=2.2$~nF.
This is done to prevent a parasitic capacitance across the inputs of $U_2$ from feeding a part of the signal from the collector of $Q_1$ to the inverting input of $U_2$, which would result in an additional pole in the loop gain, leading to a reduction of the available bandwidth.
 $R_{P6}$ and  $C_{P6}$ break the symmetry between the inputs of $U_2$, but in this frequency range $V_{CC}$ is effectively bypassed to ground by $R_L$ and $C_L=100$~nF, eliminating disturbances altogether.

\section{Loop stability and bandwidth}
\label{sec:stability}

Transistor $Q_1$ is operated at gain $G_1 = - g_m R_P$, where $g_m$ is its transconductance and \mbox{$R_P = R_{P1} \parallel R_{P2}$} is the load impedance at the collector.
We neglect poles due to stray capacitances at the collector of $Q_1$, which would be in parallel with $R_P$.
We also neglect the capacitances of $Q_1$, which are all well below 1~pF ($C_{BC}$, which is multiplied by the Miller factor, is 0.08 pF according to the datasheet).
Since $g_m = {I_C}/{V_T}$, where $I_C$ is the collector current and $V_T = {kT}/{q}$, where $k$ is the Boltzmann constant and $q$ is the elementary charge, the transconductance of $Q_1$ at a given $I_C$ is larger at 77~K by a factor $\sim 4$ with respect to room temperature.
With the values reported in table \ref{tab:components}, $G_1 = 36$ at room temperature, which becomes $G_1 = 135$ at 77~K.

\begin{figure}[t]
\centering
\includegraphics[width=0.9\textwidth]{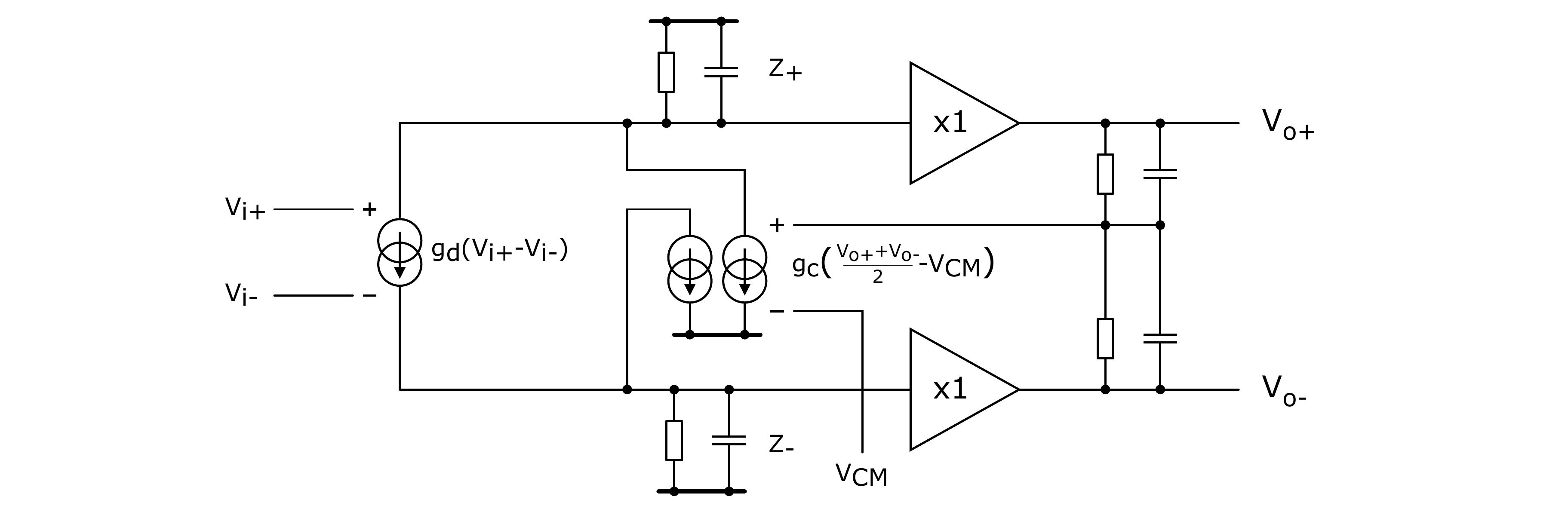}
\caption{\label{fig:schemadiffamp} Model of the fully differential operational amplifier $U_2$.}
\end{figure}

The stability of the feedback loop can be analyzed in the domain of the complex frequency~$s$.
Let us first consider the frequency response of $U_2$.
A fully differential amplifier can be modeled with the schematic shown in figure \ref{fig:schemadiffamp}.
The input stage is a voltage-controlled current generator with transconductance $g_d$.
It rejects common mode signals at the input and generates a current proportional to the differential input voltage $V_{i+} - V_{i-}$.
If we denote the load impedance of the two branches with $Z_+$ and $Z_-$, as noted in the schematic, the differential gain is given by
\begin{equation}
\frac{V_{o+}-V_{o-}}{V_{i+} - V_{i-}} = g_d  \left(Z_{+}+Z_{-}\right).
\end{equation}
while the common mode gain is given by
\begin{equation}
\frac{V_{o+}+V_{o-}}{V_{i+} - V_{i-}} =  g_d  \left(Z_{+}-Z_{-}\right).
\end{equation}
In case of perfect matching ($Z_+=Z_- = Z$) the differential gain becomes $2 g_d Z$ and the common mode gain vanishes.
The role of the common mode current generators with transconductance $g_c$ is then just to fix the output common mode voltage to be equal to $V_{CM}$.
Their bandwidth is unimportant, as long as $V_{CM}$ is constant.
If matching is not perfect, these generators also help in suppressing the residual common mode gain.

Let us assume to work in perfect matching.
The load impedance $Z$ gives the dominant pole of the opamp: we can define $A_2$ and $\tau_2$ so that $2 g_d Z = A_2/(1+s \tau_2)$.
In reality there will be at least another pole at higher frequency, for instance due to the output buffers, with time constant $\eta_2$.
The differential gain can then be modeled as
\begin{equation}
\frac{V_{o+}-V_{o-}}{V_{i+} - V_{i-}} = G_2 (s) =  \frac{A_2}{1+ s \tau_2} \frac{1}{1+ s \eta_2} \simeq  \frac{A_2}{s \tau_2} \frac{1}{1+ s \eta_2}.
\label{eq:G2definition}
\end{equation}
Assuming $U_2$ to be unity-gain stable, and not overcompensated, the frequency of the second pole corresponds to the gain-bandwidth product of the amplifier, that is the frequency where $\left| G_2 (s) \right| \simeq 1$.
The gain-bandwidth product of the THS4531 is about 27~MHz at room temperature and about 50~MHz at 77~K, which give $\eta_2 \sim 5.9$~ns and $\eta_2 \sim 3.2$~ns respectively.

\begin{figure}[t]
\centering
\includegraphics[width=\textwidth]{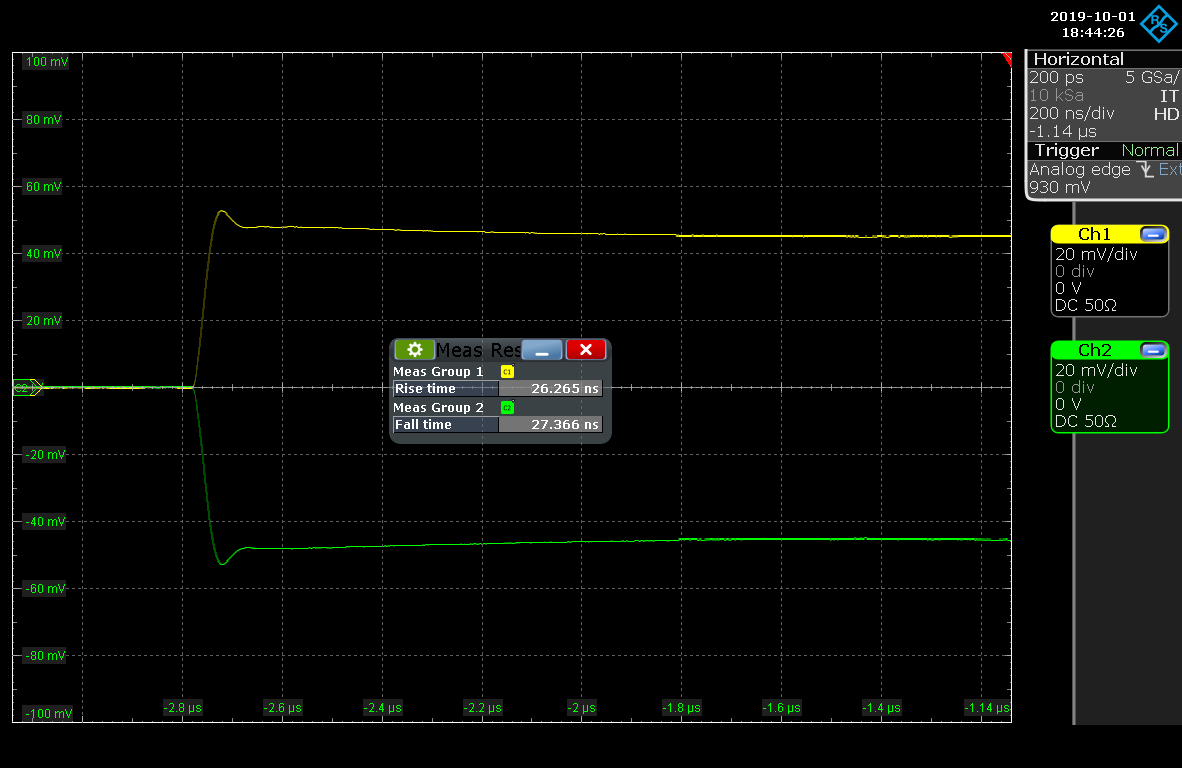}
\caption{\label{fig:outputsignals} Edges of the differential output signals for $\beta'=0.3$, driving 12~m output cables. The horizontal scale is 0.2~$\mu$s/div, the vertical scale is 20~mV/div.}
\end{figure}

Typically the feedback loop in a fully differential opamp is closed on both sides: between $V_{o+}$ and $V_{i-}$ and between $V_{o-}$ and $V_{i+}$, in a perfectly symmetrical configuration.
In the circuit described here, the feedback loop includes the input transistor $Q_1$, inherently single-ended, and therefore involves only one of the outputs.
This implies that the feedback factor is halved, and is given by
\begin{equation}
\beta(s) = \frac{g_m R_P}{2} \frac{Z_I} {Z_F+Z_I },
\end{equation}
where $Z_I = R_I + 1/s C_D$ and $Z_F = R_F \parallel 1/sC_F$.
(Note that, since $\beta(s)$ is halved, bandwidth will be halved as well, compared to the case where feedback is applied symmetrically on both branches.)
The loop gain is given by 
\begin{equation}
T (s) = - G_2 (s) \beta(s) = -   \frac{A_2}{s \tau_2}  \frac{1}{1+ s \eta_2} \left[ \frac{g_m R_P}{2}  \frac{1+s \left(C_F R_F + C_D R_I\right) + s^2 C_F R_F C_D R_I }{s ( C_D R_F+ C_D R_I + C_F R_F)+ s^2 C_F R_F C_D R_I} \right].
\end{equation}
It is clear that if both $R_I$ and $C_F$ are zero, the last term reduces to a pole at the origin, and the loop gain is unstable.
Either $R_I$ or $C_F$ need to be present to compensate one of the poles.
At the same time, as already discussed, their value should not be too large to avoid integration of the SiPM signals.
The presence of both $C_F$ and $R_I$ gives the terms in $s^2$, that become dominant at high frequency.
In principle, the presence of $R_I$ is sufficient: indeed in the measurements presented here, $C_F$ was not used, see table \ref{tab:components}.
Setting $C_F = 0$ simplifies the expression to
\begin{equation}
T (s) = -   \frac{A_2}{s \tau_2}  \frac{1}{1+ s \eta_2} \left[\frac{g_m R_P }{2} \frac{1+s C_D R_I}{s C_D \left(R_F+ R_I \right)}\right].
\label{eq:loopgain}
\end{equation}
The expression has three poles and one zero.
If $C_D R_I > \eta_2$, which is true with the choice of components in table \ref{tab:components} and $C_D$ above 10~nF, the zero compensates the pole due to $C_D (R_F+R_I)$ before the frequency of the third pole is reached.
For smaller values of $C_D$, the value of $R_I$ needs to be increased, for instance to 5.1~$\Omega$.
The condition for stability is then for the magnitude of the loop to be below unity before reaching the frequency of the pole due to $\eta_2$, which is the bandwidth limit of $U_2$.
Since in eq. \ref{eq:G2definition} we assumed $\left| G_2 (s=1/\eta_2) \right| \simeq 1$, and using the fact that $R_F \gg R_I$, this condition is satisfied if
\begin{equation}
\beta' = \frac{ g_m R_P}{2} \frac{R_I}{R_F} < 1.
\label{eq:loopstab}
\end{equation}
In other words, the amplifier is stable if the attenuation of the signal fed from ouput of $U_2$ back to the input of $Q_1$ is larger than the gain $|G_1| = g_m R_P$ provided by the first stage.
The bandwidth of the full circuit is the frequency where $\left| T(s) \right| = 1$, and coincides with the bandwidth of $U_2$ multiplied by $\beta'$.
Or, to see it differently, the gain-bandwidth product of the gain block composed by $Q_1$ and $U_2$ is $G_1$ times the gain-bandwidth product of $U_2$, and gives 6.7 GHz at 77~K.
The closed loop bandwidth of the circuit is then obtained by dividing this by $2 R_F/R_I$.
With the values of table \ref{tab:components}, eq. \ref{eq:loopstab} is well satisfied, since $\beta'=1/18$ at 77~K.
The validity of the calculations above was checked by increasing the value of $R_I$ to $5.1~\Omega$.
In this case $\beta' = 0.3$. Since the gain-bandwidth product of the THS4531 at 77~K was measured to be close to 50~MHz, the bandwidth of the full circuit with $R_I = 5.1$~$\Omega$ is expected to be about 15 MHz.
A 10\% to 90\% risetime of 23~ns was indeed observed with 2 m long output cables, in good agreement with the expected bandwidth.
If the length of the output cables is increased to 12~m, the edges of the two outputs are just slightly deteriorated to $26-27$~ns, as shown in figure \ref{fig:outputsignals}.

There is another possible source of instability of the circuit, which is not related with the entire feedback loop discussed above, but rather with self-oscillations of $Q_1$. 
Silicon-germanium HBTs are designed for radiofrequency applications, and they exhibit bandwidth in the tens or hundreds of GHz when they are biased with typical currents of tens of mA.
When biased at lower currents, as in our case, the bandwidth reduces but is still in the GHz range.
Parasitic inductance and capacitance in the layout of the circuit board can introduce unwanted resonances, which can become critical at cryogenic temperature, where the transconductance is largest.
A parasitic oscillation of $Q_1$ can appear at the output of the circuit as an oscillation at lower frequency, due to the interplay between the high frequency oscillation of $Q_1$ and the rest of the circuit.
To avoid this, the following measures were found to be effective: minimizing the stray inductance at the emitter, by connecting it to the ground plane of the board with very short traces and several vias; adding a small resistor in series with the base, a role that is already filled by $R_I$; adding the series combination of a capacitor ($\sim$nF) and resistor (a few $\Omega$) between the base and ground, close to the base, which are $C_B$ and $R_B$ in the schematic of figure \ref{fig:schematic}.
$R_I$, $R_B$ and $C_B$ can also help in achieving loop stability in case the SiPM or its connecting wires have an inductive component, which would appear in series with $C_D$.
Although they are foreseen in the schematic of figure \ref{fig:schematic}, $R_B$ and $C_B$ were not always used, and do not appear in table \ref{tab:components}.


\section{Noise}
\label{sec:noise}

\begin{figure}[t]
\centering
\includegraphics[width=\textwidth]{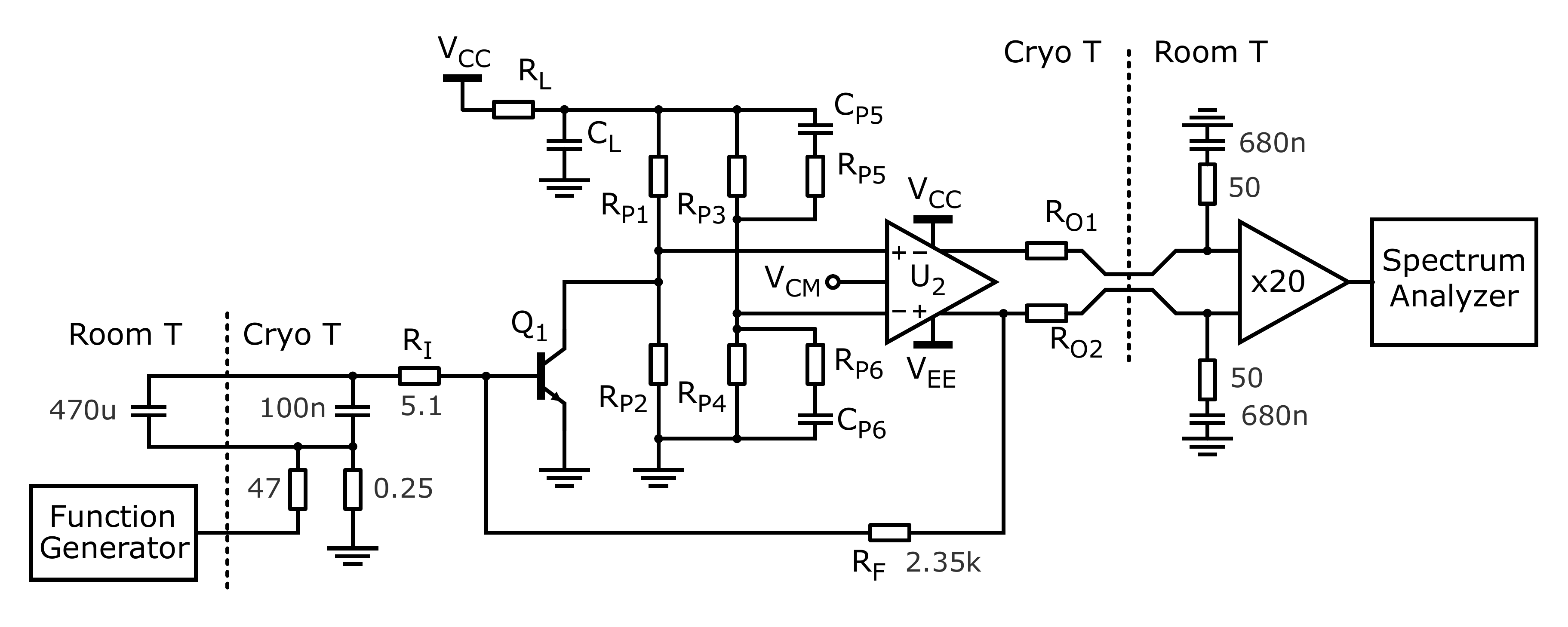}
\caption{\label{fig:schemanoise} Setup for noise measurements. All component values are those in table \ref{tab:components}, except where specified. The bias current of $Q_1$ was changed by using $R_{P1}=6.8$~k$\Omega$, $4.7$~k$\Omega$, $3$~k$\Omega$, $2.4$~k$\Omega$, giving $I_C=$~120, 210, 370, 480~$\mu A$.}
\end{figure}

The noise of $U_2$ is referred to the base of $Q_1$ by dividing it by the gain of the first stage $G_1$.
With the component values shown in table \ref{tab:components}, $G_1 = 36$ at room temperature and $G_1 =135$ at 77~K, making the noise of $U_2$ negligible.
The series noise of the amplifier is then uniquely determined by $Q_1$ and $R_I$.
The white component is given by
\begin{equation}
e_n^2 = \frac{2 k T}{g_m} + 4 k T R_{BB'} + 4 k T R_{I},
\label{eq:BJTnoise1}
\end{equation}
where $k$ is the Boltzmann constant, $T$ is the temperature, $g_m$ and $R_{BB'}$ are the transconductance and the base spreading resistance of $Q_1$.
The first and last term of eq. \ref{eq:BJTnoise1} can easily be calculated, while the second is harder to determine.
The value of $R_{BB'}$ for a given HBT at room temperature, if not explicitly listed in the datasheet, can be extracted from the quoted noise figure, and is typically a few~$\Omega$ at currents of 1~mA and above.
Its value, however, may be different for currents in the hundred $\mu$A range, and may depend on temperature.
We accounted for such dependence with a first order expansion in powers of $1/I_C$:
\begin{equation}
R_{BB'} (I_C) = R_{BB'}^{*} + \frac{\alpha}{I_{C}}
\label{eq:RBBform}
\end{equation}
where $R_{BB'}^{*}$ is the value at high collector currents, and $\alpha$ accounts for a possible increase of $R_{BB'}$ at low values of $I_C$.
Both $R_{BB'}^{*}$ and $\alpha$ can be expected to depend on temperature.
By using \ref{eq:RBBform}, and since $g_m = q I_C /kT$, where $q$ is the elementary charge, eq. \ref{eq:BJTnoise1} can be rearranged as
\begin{equation}
e_n^2 = \left(\frac{2 k^2 T^2}{q} + 4 k T \alpha \right) \frac{1}{I_C} + 4 k T \left( R_{BB'}^{*} +  R_{I} \right).
\label{eq:BJTnoise2}
\end{equation}
The sum is composed of four terms, grouped in pairs.
Measurements at different values of $I_C$ allow to disentangle the first and second from the third and fourth.
The first term, coming from the $g_m$ of $Q_1$, has the strongest temperature dependence.
We will show that, in our measurements, the first term dominates over the second at room temperature.
At cryogenic temperature, the relative importance of the second term becomes larger, and we used eq. \ref{eq:BJTnoise2} to determine the values of $R_{BB'}^{*}$ and $\alpha$ from the measured noise spectra.


The low frequency noise of transistors at cryogenic temperature depends strongly on the presence of impurities in the band gap, and, being $Q_1$ a heterojunction device, at the interface between the different semiconductors.
It is device-dependent, and possibly also batch-dependent.
The relative weight of the low frequency contribution in affecting the signal-to-noise ratio depends on the bandwidth of interest.
For a typical application that needs to detect scintillation light, the lowest frequency of interest can be associated with the scintillation time constant, a few $\mu$s in the case of liquid argon, corresponding to a few tens of kHz.
The noise contributions below the lowest frequency of interest can often be filtered out, and their weight reduced.

The current (parallel) noise is due to the base current of $Q_1$ and to the thermal noise of $R_F$.
The base current of $Q_1$ is expected below a few $\mu$A, and $R_F = 1.2$~k$\Omega$.
Any base current below 100~$\mu$A and feedback resistors above 100~$\Omega$ give a white current noise below 10~\pAHz.
For large values of $C_D$, and for signal frequencies above the kHz range, this contribution is negligible.

\begin{figure}[t]
\centering
\includegraphics[width=0.75\textwidth]{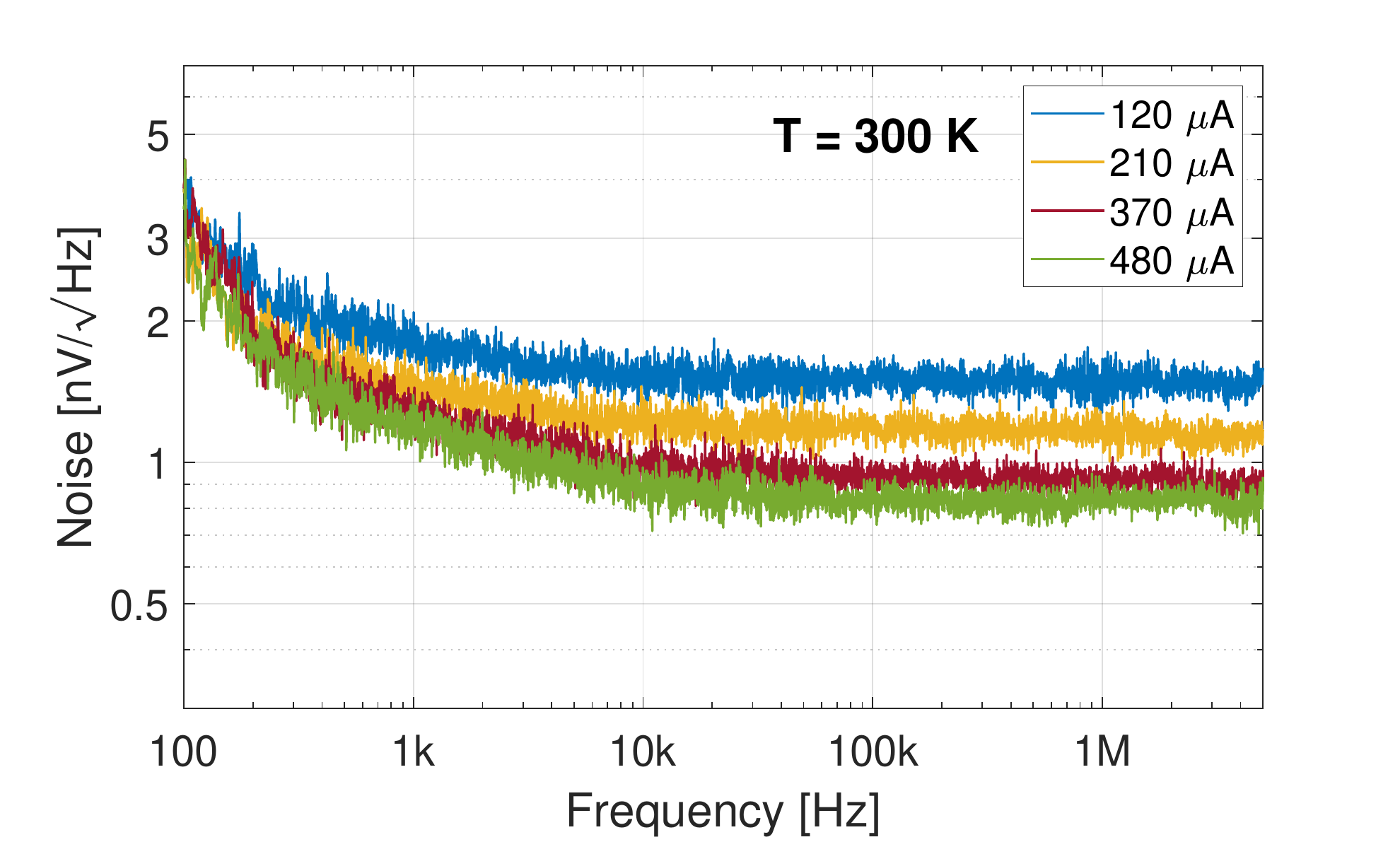}
\centering
\includegraphics[width=0.75\textwidth]{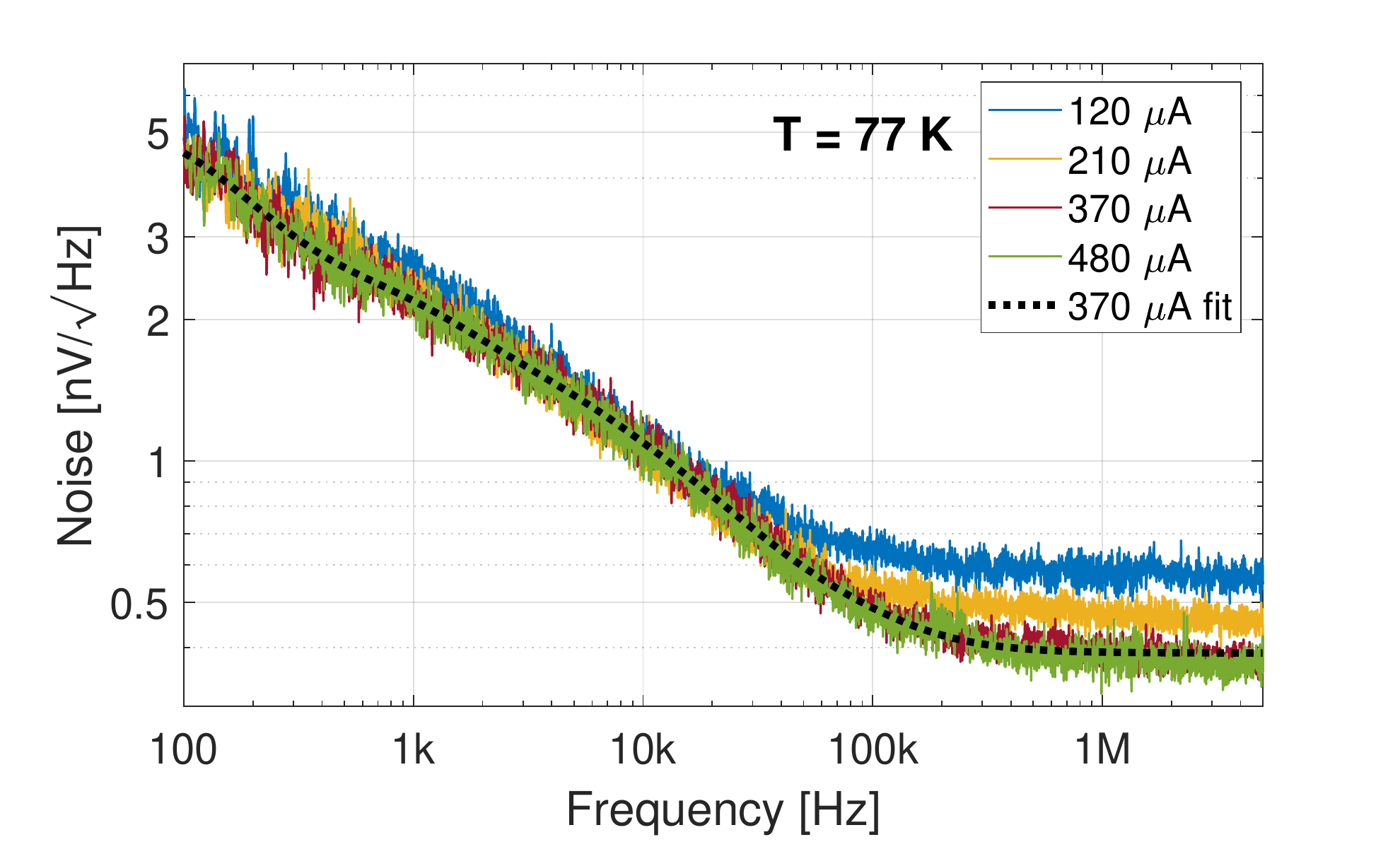}
\caption{\label{fig:noisespectra}Noise spectra measured for different values of the collector current $I_C$ at room temperature (top) and liquid nitrogen (bottom).
At 77~K, the spectra at 370~$\mu$A and 480~$\mu$A are almost indistinguishable.
The black dotted curve shows the interpolation of the data at 77~K, 370~$\mu$A.}
\end{figure}

The noise spectra were measured in the configuration shown in figure \ref{fig:schemanoise}.
The amplifier was operated at a differential gain close to $900$ ($R_F = 2.35$~k$\Omega$, $R_I = 5.1$~$\Omega$).
The input node, before $R_I$, was connected to ground through the series connection of a small 0.25~$\Omega$ resistor and a large capacitance, composed of a $470~\mu$F tantalum capacitor at room temperature, connected to the input with a 20 cm cable, in parallel with ~100 nF C0G and 10 $\mu$F X7R ceramic capacitors at cold.
The differential output of the circuit was converted to single-ended on a second stage amplifier based on a AD8055 with gain 20, closed-loop bandwidth 10~MHz, and measured with a Rohde\&Schwarz FSV4 spectrum analyzer.
To characterize the transfer function, a test signal from an Agilent 33250A (white noise, 100~mV peak to peak, bandwidth 80~MHz) was injected through a $50+47$~$\Omega$ source impedance across the 0.25~$\Omega$ resistor, resulting in a 0.25~mV peak to peak signal at the input node of the circuit.
The source was then disconnected to measure the output noise spectrum, which was then divided by the measured transfer function.

Figure \ref{fig:noisespectra} shows the results at room temperature (300~K) and liquid nitrogen (77~K) for different values of the collector current $I_C$.
(Each spectrum was obtained by concatenating several measurements taken on different time scales.)
The white noise depends on $I_C$ as expected.
At larger values of $I_C$ the improvement becomes less evident, due to the presence of the constant terms in eq. \ref{eq:BJTnoise2}.
The low frequency part of the spectra does not depend on $I_C$, and is slightly more pronounced at 77~K.
The curves show lorentzian ``bounces'' below $10$~kHz.
The squared spectra were interpolated with the sum of five lorentzian functions and a constant (white) term:
\begin{equation}
N(\omega) = A(\omega) + e_n^2 =  \sum_{i=1}^5 \frac{A_i}{1+\omega^2 \tau_i^2}   +  e_n^2 
\label{eq:noiseinterp}
\end{equation}
Each lorentzian term is expected to be due to the presence of a specific class of generation-recombination centers (traps) with a definite time constant $\tau_i$ \cite{Lauritzen1968,VanVliet1976}.
The most prominent are found at $\tau_i=100$~$\mu$s, corresponding to 1.5~kHz, and at about 100~Hz, accounting for the rise at the lower end of the spectrum.
The resulting curve for the spectrum at 370~$\mu$A is shown with a dotted black line in figure \ref{fig:noisespectra}.



\begin{figure}[t]
\centering
\includegraphics[width=0.8\textwidth]{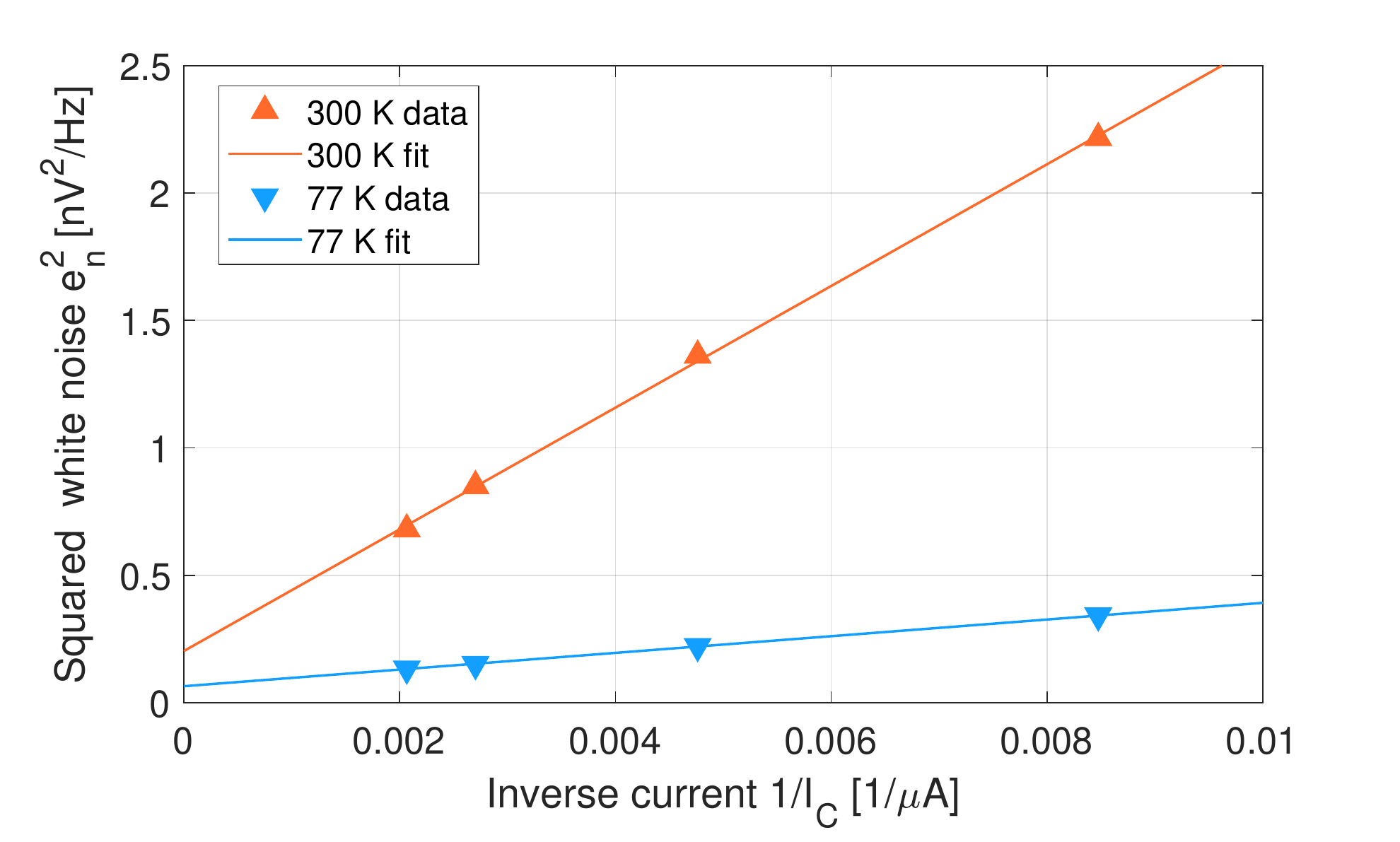}
\caption{\label{fig:whitenoisevsIC}Squared white noise extracted from the spectra of figure \ref{fig:noisespectra}, plotted as a function of the inverse collector current of $Q_1$.}
\end{figure}

From the fitting curves, the value of the white component $e_n^2$ was extracted.
Figure \ref{fig:whitenoisevsIC} gives the squared white noise $e_n^2$ as a function of $1/I_C$.
From eq. \ref{eq:BJTnoise2}, a linear fit of the data allows to extract the unknown parameters $R_{BB'}^*$ and $\alpha$.
The intercept corresponds to the condition $1/I_C \rightarrow 0$, where the first two terms in eq. \ref{eq:BJTnoise2} vanish.
From the intercept values, which are 0.203~nV$^2$/Hz at 300~K and 0.064~nV$^2$/Hz at 77~K, the values of $R_{BB'}^* + R_I$ can be determined.
Knowing that $R_I=5.1\ \Omega$, we obtain $R_{BB'}^* \sim 7\ \Omega$ at 300~K and $R_{BB'}^* \sim 10\ \Omega$ at 77~K.
The value at 300~K is compatible with the value extracted from the noise figure in the BFP640 datasheet.

The slope of the fitting line at 300~K is 239~$\mu$A(nV)$^2$/Hz.
It is about 10\% larger than $2 k^2 T^2 / q =$~214~$\mu$A(nV)$^2$/Hz.
About half of this excess noise at room temperature can be explained by the noise of the second stage $U_2$: the THS4531 has a white voltage noise of 10~\nVHz~ at room temperature, which divided by $G_1=36$ gives about 0.3~\nVHz~at the input.
The contribution of the second term of eq. \ref{eq:BJTnoise2} at 300~K is then small, and does not allow a precise determination of $\alpha$.
In other words, the value of $R_{BB'}$ at room temperature is small enough not to give sizeable contributions to the total noise of the circuit.

At 77~K, the slope of the fitting curve gives 32.7~$\mu$A(nV)$^2$/Hz at 77~K, which is twice as large as what is expected from the first term of eq. \ref{eq:BJTnoise2} alone, $2 k^2 T^2 / q =$~14.1~$\mu$A(nV)$^2$/Hz.
The noise of the THS4531 was observed to decrease to about 3.3~\nVHz~ at 77~K, while the gain of the first stage increases to $G_1 = 135$ at 77~K, making it completely negligible.
Since the circuit was submerged in liquid nitrogen, and the power consumption is low, we ruled out the possibility that $Q_1$ was operating at a temperature significantly higher than 77~K.
The difference gives $\alpha = 4380\ \mu A \Omega$.
The estimated value of $R_{BB'}$ is then 18, 22, 30, 46 $\Omega$ at $I_C=$~480, 370, 210, 120~$\mu A$ respectively.

Even though partially contributed by $R_{BB'}$, the total white noise of the amplifier is remarkably low, considered the low power consumption.
In all the measurements presented in the following sections, we worked at $370\ \mu$A, with a white noise of 0.4~\nVHz.
As can be clearly seen in figure \ref{fig:noisespectra}, working with larger currents than those considered did not give any advantage at 77~K.

\section{Performance in reading out SiPMs}

\begin{figure}[t]
\centering
\includegraphics[width=0.75\textwidth]{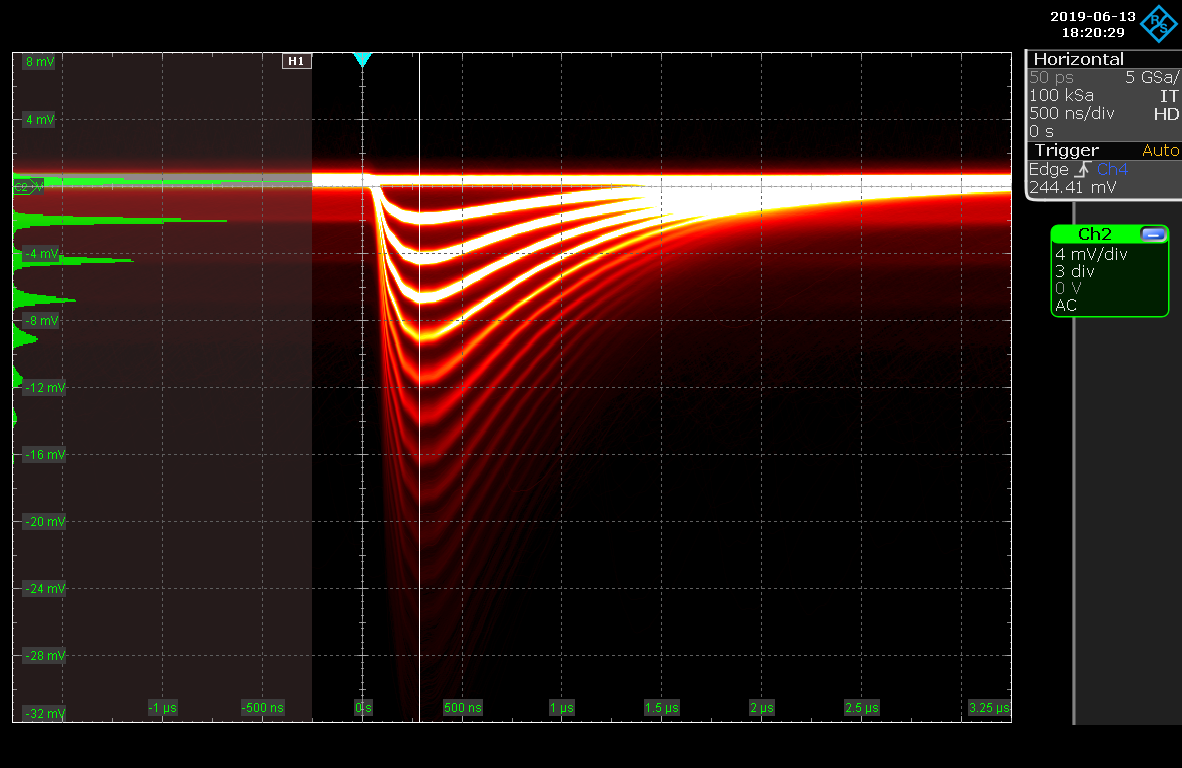}
\caption{\label{fig:FBKtile}Light signals from a pulsed LED, detected by a SiPM, read out with the amplifier described in this paper. The horizontal scale is 500~ns/div, the vertical scale is 4~mV/div.}
\end{figure}

Given the measured noise spectra, we address the determination of the signal-to-noise ratio in the readout of actual SiPMs.
Figure \ref{fig:FBKtile} shows an example of a SiPM read out at 77~K (liquid nitrogen) with the amplifier described in this paper.
The SiPM has an area of 0.96~cm$^2$, a total capacitance of 4.8~nF, and a recovery time $\tau_D \simeq 800$~ns.
It was biased at 24~V (3~V overvoltage) by setting the cathode voltage to 25~V, since the anode, connected to the input of the circuit, was at 1~V.
The gain of the SiPM at 3~V overvoltage was about $2.4\times10^6$.
It was illuminated with a pulsed LED through an optical fiber, and the light intensity was adjusted so that a few photoelectrons were detected on each pulse.
Figure \ref{fig:FBKtile} shows excellent separation between signals corresponding to different numbers of photoelectrons.

\begin{figure}[t]
\centering
\includegraphics[width=0.8\textwidth]{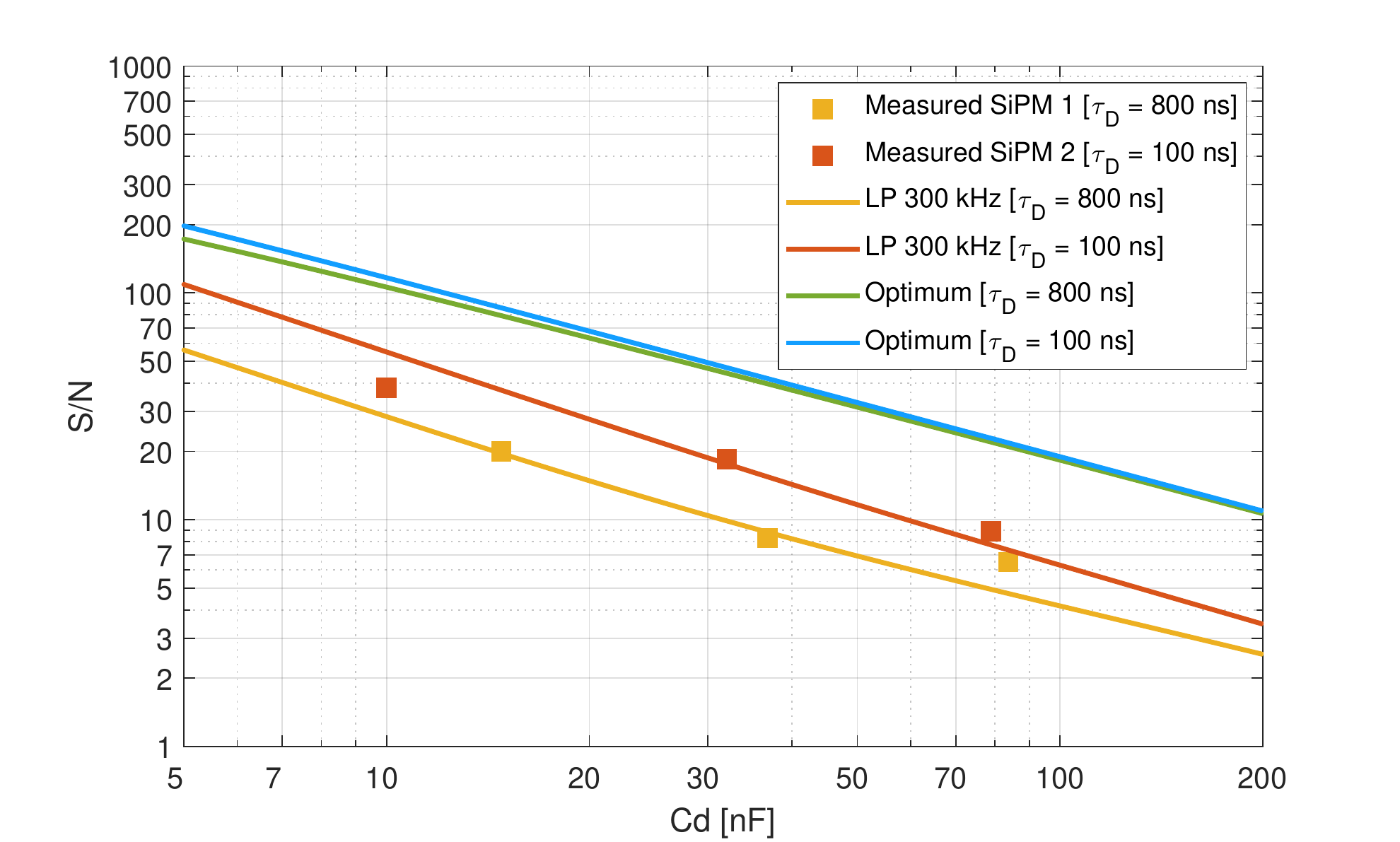}
\caption{\label{fig:SNcomparison}Signal-to-noise ratio for single photoelectron signals from two SiPMs operated at a gain of $2.4\times 10^6$. The SiPMs differ in the values of the recovery time $\tau_D$.
The measured data at different capacitance values for two devices, labelled ``Measured SiPM 1'' and ``Measured SiPM 2'', are in reasonable agreement with the expected curves ``LP 300 kHz'', calculated from the gain and $\tau_D$ of the devices and the measured noise spectra in \ref{fig:noisespectra}, using just a low pass filter at 300~kHz.
The curves can also be compared with the expected ``Optimum'' $S/N$, obtained by integrating \ref{eq:optimum2} numerically.}
\end{figure}

To determine the signal-to-noise ratio in the case of larger photosensitive area, we added different values of capacitance at the input node: 10~nF, 32~nF and 79~nF.
Considering as reference a SiPM capacitance of 50~pF/mm$^2$, these values correspond to roughly 2~cm$^2$, 6.4~cm$^2$ and 15.8~cm$^2$ respectively, to be added to the 0.96~cm$^2$ of the device.
For the first point at 10~nF, the value of $R_I$ was increased from 1~$\Omega$ to 5.1~$\Omega$ to guarantee stability of the loop gain.
A warm second stage based on a OP27 opamp, with bandwidth 300~kHz and gain 23.5, was used to convert the differential signals to single ended.
The bandwidth was further filtered at 1~MHz on the oscilloscope.
The signal-to-noise ($S/N$) for single photoelectron signals was determined by dividing the observed signal amplitude by the RMS noise of the baseline.
The resulting data points are shown in figure \ref{fig:SNcomparison} as ``Measured SiPM 1''.
The data are in reasonable agreement with the expected behaviour, which is described by the curve labelled ``LP 300 kHz [$\tau_D=800$~ns]''.
The curve was obtained by calculating the expected signal amplitude for single photoelectron signals from a SiPM with gain $2.4\times 10^6$, $\tau_D=800$~ns, filtered with a single-pole low-pass at 300~kHz, and dividing it by the RMS noise obtained by integrating numerically from 100~Hz to 1~MHz the noise spectrum of the amplifier, low-pass filtered at 300~kHz. 

The same measurement was repeated for a different device, with an area of 1.7~mm$^2$, a capacitance of 35~pF, and a shorter recovery time $\tau_D \simeq 100$~ns.
The gain of the device at 3~V overvoltage is expected to be about $1.7\times 10^6$, hence it was operated at a slightly higher overvoltage, adjusted to obtain a gain close to $2.4\times 10^6$.
As done in the previous case, capacitances of 10~nF, 32~nF and 79~nF were added in parallel with the SiPM to simulate a larger total area.
The measured $S/N$ is shown in figure \ref{fig:SNcomparison} as ``Measured SiPM 2''.
Again, the data are in reasonable agreement with the expected values, calculated for a 300~kHz low pass filter, labelled as ``LP 300~kHz [$\tau_D=100$~ns]''.

By comparing the two sets of measurements, it appears that a shorter recovery time $\tau_D$ gives a better $S/N$ by almost a factor two, although the difference becomes smaller at higher values of $C_D$.
This is due to the larger weight of the low frequency noise components for larger values of $\tau_D$.
In many applications, however, a relatively long recovery time is beneficial for the suppression of afterpulses.
In any case, even at larger values of $\tau_D$, the measurements show a good $S/N$, always above 4 up to 100~nF capacitance, which would correspond to a photosensitive surface of 20~cm$^2$ for a 50~pF/mm$^2$ SiPMs.

It could be argued that the 300~kHz low pass filter applied in these measurements is not the optimal one, and the $S/N$ could improve with better filtering algorithms, to be applied offline.
From the theory of optimal filtering of signals from particle detectors \cite{Radeka1967,Gatti1986,Gatti2004}, the best signal-to-noise ratio that can be achieved for any amplitude measurement is given by (``OF'' stands for optimum filter):
\begin{equation}
\left(\frac{S}{N}\right)^2 _{\textrm{OF}} = \frac{1}{2 \pi} \int_{-\infty} ^{\infty} \frac{\left|{\tilde{V}_O}(\omega)\right|^2}{N(\omega)}  d \omega
\label{eq:optimum1}
\end{equation}
where ${\tilde{V}_O(\omega)}$ is the Fourier transform of the output signal $V_{O}(t)$, and $N(\omega)$ is the output noise spectral density.
Any other filter or processing algorithm is bound to give a sub-optimal signal-to-noise ratio.
The signal of a single SiPM cell, corresponding to a single photoelectron, can be described by a current step with an exponential decay, carrying a total charge $Q$.
The differential signal at the output of the amplifier can then be expressed as
\begin{equation}
V_O(t) = 2 R_F I_D (t) = 2 R_F \frac{Q}{\tau_D} \textrm{e}^{-\frac{t}{\tau_D}}.
\label{eq:outputsignal123}
\end{equation}
To match the gain of the SiPMs we measured, we set $Q = 2.4 \times 10^6$~$q$, where $q$ is the elementary charge, and we consider two values for $\tau_D$, 100~ns and 800~ns.
Here and in the following we neglect the integration at the input node, by considering $R_I=0$, and neglect the finite bandwidth of the amplifier.
(Even if we wanted to consider these effects in eq. \ref{eq:outputsignal123}, they would affect the following expression \ref{eq:outputnoise123} for the output noise in the same way.
They would therefore cancel out from the integrand of eq. \ref{eq:optimum2}.)
The magnitude of the Fourier transform of the output signal is
\begin{equation}
\left|\tilde{V}_O(\omega)\right|^2 = \left|  \frac{2 R_F Q}{1+i\omega \tau_D}\right|^2 =  \frac{4 R_F^2 Q^2}{1+\omega^2 \tau_D^2}.
\end{equation}
The output noise spectral density, neglecting parallel contributions, is given by
\begin{displaymath}
N_O(\omega) = \left| \frac{2 \left(Z_F+Z_I \right)}{Z_I}\right|^2  \left(A(\omega) + e_n^2\right) =   \left| 2+2 i\omega C_D R_F \right|^2 \left(A(\omega) + e_n^2\right)
\end{displaymath}
\begin{equation}
= \left(4+4\omega^2 C_D^2 R_F^2 \right) \left(A(\omega) + e_n^2\right)
\label{eq:outputnoise123}
\end{equation}
where we used $Z_F=R_F$ and $Z_I = 1/s C_D$.
The terms $A(\omega)$ and  $e_n^2$ account for the low frequency and white noise respectively, as expressed by eq. \ref{eq:noiseinterp}.
The integral in eq. \ref{eq:optimum1} becomes
\begin{equation}
\left(\frac{S}{N}\right)^2 _{\textrm{OF}} = \frac{1}{2 \pi} \int_{-\infty} ^{\infty}  \frac{R_F^2 Q^2}{\left(1+\omega^2 \tau_D^2\right) \left(1+\omega^2 C_D^2 R_F^2\right) \left(A(\omega) + e_n^2\right)} d \omega
\label{eq:optimum2}
\end{equation}
Neglecting the low frequency noise allows to calculate the integral analytically, but leads to an overestimated result.
We can instead take the actual noise spectrum, as shown in figure \ref{fig:noisespectra}, and evaluate the integral numerically.
The resulting curves are shown in figure \ref{fig:SNcomparison}, labelled as ``Optimum'' for the two values of $\tau_D$.
The dependence of the curves on $\tau_D$ is small at low values of $C_D$, and completely negligible for larger values of $C_D$.
The curves show that, with respect to the measured $S/N$ discussed previously, there is margin for improvement by up to factors of $3-4$ if better processing algorithms than a simple 300~kHz low pass filter were applied.
In this case, a $S/N$ above 10 would be expected for SiPM capacitance up to 200~nF, corresponding to approximately 40~cm$^2$, independently of the SiPM recovery time.

\section{Conclusions}

We presented a front-end amplifier designed to readout large arrays of SiPMs in cryogenic environments.
Compared with similar amplifiers based only on operational amplifiers, the present design gives lower noise at significantly lower power dissipation.
The resulting gain-bandwidth product is in the GHz range.
The circuit topology offers high flexibility in compensating the loop gain, which allows to obtain a signal rise time down to $\sim20$~ns with large values of input capacitance, up to $\sim100$~nF.
The conditions for close-loop stability were discussed in detail.
The circuit shows input-referred white voltage noise of 0.4~\nVHz\ at 77~K, at a power consumption of 2.5~mW.
The base spreading resistance of the input transistor was determined from white noise measurements, and was observed to depend on its bias current.
At low frequency, noise is also contributed by lorentzian terms.
The amplifier allowed to measure $S/N>4$ for single photoelectron signals with input capacitances up to 100~nF, corresponding to a total SiPM surface of about 20~cm$^2$.
Further improvements would derive from applying an optimum filtering algorithm, with the expected capability to readout a 100~nF photosensitive surface with $S/N \simeq 20$, independently of the SiPM recovery time.

\end{document}